\documentclass[aps,pra,twocolumn,10pt,letterpaper,superscriptaddress]{revtex4-1}

\usepackage{graphicx}
\usepackage{xcolor}
\usepackage{dcolumn}
\usepackage{bm}

\usepackage{mathtools}
\usepackage{amsfonts}
\usepackage{graphicx}
\usepackage{hyperref}
\newcommand{\bra}[1]{\ensuremath{\left\langle#1\right|}}
\newcommand{\ket}[1]{\ensuremath{\left|#1\right\rangle}}
\expandafter\let\csname equation*\endcsname\relax
\expandafter\let\csname endequation*\endcsname\relax
  
\usepackage{amsmath}  
\usepackage{color}
\usepackage{mathtools}
\usepackage{pgfplots}
\usepackage{graphicx}
\usepackage{braket}
\usepackage{stackengine}

\renewcommand{\epsilon}{\varepsilon}

\begin{document}

\title{Branching high-order exceptional points in non-hermitian optical systems}

\author{Konrad~Tschernig}\email{Konrad.Tschernig@mbi-berlin.de}
\affiliation{Max-Born-Institut, Max-Born-Stra\ss{}e 2A, 12489 Berlin, Germany}
\affiliation{Humboldt-Universit\"at zu Berlin, Institut f\"ur Physik, AG Theoretische Optik $\&$ Photonik, 12489 Berlin, Germany}

\author{Kurt Busch}
\affiliation{Max-Born-Institut, Max-Born-Stra\ss{}e 2A, 12489 Berlin, Germany}
\affiliation{Humboldt-Universit\"at zu Berlin, Institut f\"ur Physik, AG Theoretische Optik $\&$ Photonik, 12489 Berlin, Germany}

\author{Demetrios N. Christodoulides}
\affiliation{CREOL, The College of Optics and Photonics, University of Central Florida, Orlando, FL 32816-2700, USA}

\author{Armando Perez-Leija}\email{apleija@gmail.com}
\affiliation{Max-Born-Institut, Max-Born-Stra\ss{}e 2A, 12489 Berlin, Germany}
\affiliation{Humboldt-Universit\"at zu Berlin, Institut f\"ur Physik, AG Theoretische Optik $\&$ Photonik, 12489 Berlin, Germany}

\date{\today}

\begin{abstract}
Exceptional points are complex-valued spectral singularities that lead to a host of intriguing features such as loss-induced transparency - a counterintuitive process in which an increase in the system's overall loss can lead to enhanced transmission. In general, the associated enhancements scale with the order of the exceptional points. Consequently, it is of great interest to devise new strategies to implement realistic devices capable of exhibiting 
high-order exceptional points. 
Here, we show that high-order $N$-photon exceptional points can be generated by exciting non-hermitian waveguide arrangements with coherent light states. Using photon-number resolving detectors it then becomes possible to observe $N$-photon enhanced loss-induced transparency in the quantum realm. 
Further, we analytically show that the number-resolved dynamics occurring in the same nonconservative waveguide arrays will exhibit eigenspectral ramifications having several exceptional points associated to different sets of eigenmodes and dissipation rates.
\end{abstract}

\maketitle


\section{Introduction}
Exceptional points (EPs) are spectral degeneracies that emerge in open dynamical systems \cite{Bender_2007}. 
For dissipative systems, the associated eigenspectra are functions of the dissipation rates and an EP occurs at a critical dissipation rate $\Gamma_c$ around which the real and imaginary part of two or more eigenvalues coalesce and bifurcate, respectively \cite{Schnabel2017,Wang2019}.
Physically, the spectral coalescences and bifurcations imply that some of the eigenmodes supported by a system affected by a dissipation rate smaller than $\Gamma_c$ will evolve acquiring different phases while experiencing identical dissipation rates.
In contrast, by driving the system to operate beyond the EP, the corresponding eigenmodes will now evolve with equal phases and different dissipation rates.
Counterintuitively, these contrasting features indicate that by increasing the dissipation rate in a system designed to work around the EP, one can readily upgrade the system to support slowly decaying eigenmodes that can propagate for much longer distances. 
In practice, these peculiar effects have been exploited to magnify the otherwise negligible responses of a variety of dissipative systems \cite{Mirieaar7709}.\\

In optics, the steep eigenspectral slopes encountered in the vicinity of the EPs have lead to the development of very interesting applications, such as  loss-induced transparency \cite{SiviloglouPT}, laser mode selectivity \cite{Feng972,Hodaei975}, and unidirectional invisibility \cite{PengPT,LinPT}.
Furthermore, it has been recognized that the enhanced sensitivity to external perturbations and noise becomes more pronounced in systems exhibiting second-order EPs \cite{Mortensen18,Wiersig2014,Brandstetter,FengLiang,RotterS,PhysRevA.98.023805}. This fact has sparked a tremendous interest in the optics community to devise realistic systems capable of exhibiting EPs of even higher order, which in principle could further magnify the effects of external perturbations, leading to exceptional non-hermitian sensors \cite{Mirieaar7709}.
In this regard, several studies in classical optics have predicted the existence of high-order EPs in waveguides trimers \cite{MariaGraefe}, photonic crystals \cite{LinEPs}, and  optomechanical systems \cite{JingNori}. Additionally, in the context of parity-time symmetric systems \cite{Bender_2007} many theoretical \cite{Klaiman2008,Graefe_2008,Longhi2010} and experimental \cite{Rutter,Regensburger2012,Stone2011} investigations have demonstrated the occurrence of EPs in a variety of systems in which optical gain is easy to implement \cite{ElGanainy07,Makris2008,Musslimani2008}. \\ 

Beyond classical optics, it has been recently formulated that the photon number-resolved dynamics of semi-dissipative waveguide beam splitters, excited by $N$ indistinguishable photons will exhibit an EP of order $N+1$ \cite{quiroz2019exceptional}. 
Yet, while it is nowadays rather routine to generate nonclassical states with pairs of indistinguishable photons, the controlled generation of states involving $N>2$ photons remains challenging \cite{Silberhorn,Magana}.\\

In this work we outline a theoretical prescription to generate EPs of arbitrary order by exciting nonconservative waveguide systems with coherent states \cite{Sudarshan}. The advantages of using coherent states are twofold: firstly, coherent states are natural eigenstates of the annihilation operator, this means that under dissipative conditions the purity of coherent states remains unaffected, and secondly, coherent states can be easily generated using strongly attenuated laser light \cite{Afek879,Hofmann2007,Heilmann2016}.
The $N$-photon EPs are then exposed by projecting the coherent output states onto particular $N$-photon subspaces, something that in practice can be done using photon number resolving (PNR) detectors \cite{Lita08,Young2020,Zou2020}. 
Furthermore, we show that the number-resolved dynamics of the same nonconservative waveguide arrays, excited by coherent states and post-selected to the $N$-photon subspace, will exhibit eigenspectral ramifications leading to the appearance of several EPs that emerge due to the coalescence of different sets of $N$-photon eigenmodes at different dissipation rates.\\

\section{Results}
Consider a system consisting of $M$ evanescently coupled single-mode waveguides excited with coherent states. In such a scenario, the quantum light dynamics is described by the non-hermitian Hamiltonian
\small
\begin{equation}
\hat{H} = \sum_{m=1}^M (\beta + i \Gamma_m) \hat{a}^\dagger_m \hat{a}_m + \kappa \sum_{m=1}^{M-1} \left(\hat{a}^\dagger_m \hat{a}_{m+1} + \hat{a}^\dagger_{m+1} \hat{a}_m\right),
\end{equation}
\normalsize
where $\hat{a}^\dagger_m$ ($\hat{a}_m$) is the creation (annihilation) operator, $\beta$ is the propagation constant, $\kappa$ is the coupling coefficient between adjacent waveguides, and $\Gamma_m$ is the dissipation coefficient of the $m$'th waveguide. Throughout this work we assume that the waveguides are tuned such that they exhibit identical propagation constants $\beta=0$. Note, we exclusively consider lossy systems $(\Gamma_m\geq0)$, which are far easier to implement than systems with gain, $(\Gamma_m<0)$. \\

We start by examining the simplest configuration consisting of a semi-lossy waveguide beam splitter with $\Gamma_1=\Gamma$ and $\Gamma_2=0$. This particular arrangement features a second order EP. That is, the two eigenvalues of the corresponding Hamiltonian, $\lambda_{\pm}=i\Gamma/2 \pm \sqrt{\kappa^2-\Gamma^2/4}$, become degenerate at the critical dissipation rate $\Gamma_c=2\kappa$. Concurrently, the eigenmodes $\ket{\phi_\pm}=\ket{\alpha\lambda_\pm/(\sqrt{2}\kappa)}_1 \ket{\alpha/\sqrt{2}}_2$ coalesce into the \textit{exceptional mode} (EM) $\ket{\phi_+}=\ket{\phi_-}=\ket{i\alpha/\sqrt{2}}_1\ket{\alpha/\sqrt{2}}_2 \equiv \ket{\phi_E}$. 
Note the eigenmodes have been expressed in terms of coherent states $\ket{\alpha}$ \cite{Sudarshan}, and the subscripts $1$, $2$ indicate waveguide labels. It is worthwhile to mention that the EM can be easily generated by exciting any port of a lossless 50/50 waveguide beamsplitter with a coherent state $\ket{\alpha}$.
Further, by adding a small perturbation to the critical dissipation coefficient, $\Gamma=\Gamma_c+\epsilon$, reveals the typical $\sqrt{\epsilon}$-scaling of the differences of the eigenvalues $\lambda_+-\lambda_-=2i\sqrt{\kappa}\sqrt{\epsilon}$ at the EP \cite{HodaeiNature}.

The response of this semi-lossy waveguide beam splitter at the critical $\left(\Gamma=2\kappa\right)$ and super-critical $\left(\Gamma>2\kappa\right)$ regimes is computed analytically using the evolution matrices
\small
\begin{equation}
\hat{U}_c (z) = e^{-\frac{\Gamma z}{2}}\begin{pmatrix}
1- \kappa z& i\kappa z\\
i \kappa z& 1 + \kappa z
\end{pmatrix},
\end{equation}
\normalsize
and
\small
\begin{equation}\label{eq:sc}
\hat{U}_{sc} (z)= e^{-\frac{\Gamma z}{2}} \left[\frac{\sinh(\theta z)}{2\theta} \begin{pmatrix}
-\Gamma &  i2 \kappa \\
i 2 \kappa & \Gamma
\end{pmatrix} + \cosh(\theta z)\mathbb{I}
\right],
\end{equation}
\normalsize
\begin{figure*}[t!]
\includegraphics[width=0.8\linewidth]{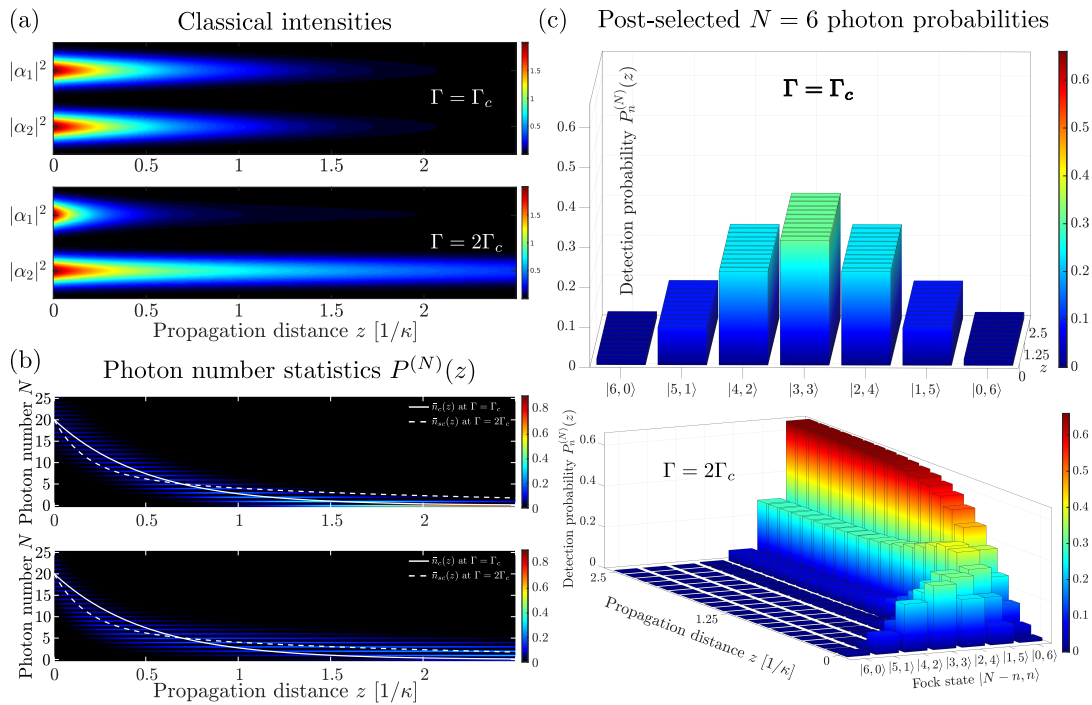}
\caption{Evolution of the exceptional mode $\ket{\phi_E}=\ket{i\alpha/\sqrt{2}}\ket{\alpha/\sqrt{2}}$ ($\alpha=\sqrt{20}$) in the semi-lossy waveguide beam splitter. \textbf{(a)} Classical intensities $|\alpha_1|^2, |\alpha_2|^2$. \textbf{(b)} Total photon number distribution $P^{(N)}$. \textbf{(c)} Probabilities to detect $n$ photons in the neutral waveguide and $N-n$ photons in the lossy one after post-selection to the $N=6$ photon subspace. In \textbf{(a)}, \textbf{(b)} and \textbf{(c)} the top and bottom panels correspond to the critical ($\Gamma=\Gamma_c$) and super-critical ($\Gamma=2\Gamma_c$) regimes, respectively. In all cases we choose $\kappa=1$, implying a critical loss rate $\Gamma_c=2$. In \textbf{(a)} we observe the classical effect of the loss-induced transparency, which manifests in \textbf{(b)} as an increase in the total number of photons after a certain propagation distance in the super-critical regime (dashed line in \textbf{(b)}) with respect to the critical point (solid line in \textbf{(b)}). Initially, the post-selected distribution \textbf{(c)} is symmetric, with the state $|3,3\rangle$ exhibiting the highest probability $\approx 30 \%$. The distribution $P_n^{(N)}(z)$ remains constant along evolution, when $\Gamma=\Gamma_c$. In the super-critical regime $P_n^{(N)}(z)$ evolves into an asymmetric distribution, where significantly more photons are detected in the neutral waveguide.}
\label{fig:exceptionalmode}
\end{figure*}
respectively. Here, we have defined $\theta = \sqrt{\Gamma^2-4 \kappa^2}/2$, and $\mathbb{I}$ is the $2\times2$ identity matrix.
Hence, in a system operating at the EP that is excited by the exceptional eigenmode $\ket{\phi_E}=\ket{i\alpha/\sqrt{2}}_1\ket{\alpha/\sqrt{2}}_2$, the state will evolve exhibiting a global damping $\ket{\psi(z)}=\hat{U}_c \ket{\phi_E}=e^{-\frac{\Gamma z}{2}}\ket{\phi_E}$. In contrast, when the system operates in the super-critical regime the same excitation leads to the state $\ket{\alpha_{1}(z)}_{1}\ket{\alpha_{2}(z)}_{2}$ with amplitudes given as $\alpha_{1}(z)=i\alpha/\sqrt{2} e^{\frac{-\Gamma z}{2}}\left(\frac{\sinh(\theta z)}{2\theta}(2\kappa-\Gamma)+\cosh(\theta z)\right)$ and $\alpha_{2}(z)=\alpha/\sqrt{2} e^{\frac{-\Gamma z}{2}}\left(\frac{\sinh(\theta z)}{2\theta}(-2\kappa+\Gamma)+\cosh(\theta z)\right)$, and average photon numbers $I_1=|\alpha_1(z)|^2$ and $I_2=|\alpha_2(z)|^2$ in the respective waveguides.
Fig.~(\ref{fig:exceptionalmode}-a) shows the analytically computed intensity evolution along the waveguides. Clearly, 
the fingerprints of the loss-induced transparency effect are observed in the intensity evolution \cite{SiviloglouPT}. 
To be precise, when the system operates at the critical point ($\kappa=1,\Gamma=2$), after a propagation distance of $z=1/\kappa$ we find the total intensity $I_c=I_1+I_2\approx 0.1348 \ I_0$ - where $I_0$ is the total input intensity - while in the super-critical regime, ($\kappa=1,\Gamma=4$), $I_{sc}\approx 0.2011 \ I_0$. That is, despite the overall increase in the dissipation rate, in the super-critical case the output light becomes brighter by a factor of $\approx 1.5$. \\

We now examine the photon statistics of this semi-lossy waveguide beam splitter. 
To do so, we express the input state $\ket{\phi_E}$ as a superposition of states with definite photon number $N$, that is, $\ket{i\alpha/\sqrt{2}}_{1}\ket{\alpha/\sqrt{2}}_{2}=\sum_{N=0}^\infty \ket{\psi^{(N)}}$, where $\ket{\psi^{(N)}}=\sum_{n=0}^N c_n^{(N)} \ket{N-n}_{1}\ket{n}_{2}$ and 
$c_n^{(N)}=\braket{N-n|i\alpha/\sqrt{2}}\braket{n|\alpha/\sqrt{2}}=e^{-|\alpha|^2/2}(\alpha/\sqrt{2})^N i^{N-n}/\sqrt{(N-n)!n!}$.
From this multiphoton picture, it is evident that the coherent EM readily excites an infinite number of $N$-photon subspaces simultaneously, and the detection of $N$-photons at the output ports projects the system onto one of the subspaces. 
Explicitly, the probability of finding the system in the $N$-photon subspace, which is equivalent
to the probability of detecting $(N-n)$ photons at port 1 and $n$ photons at port 2, is $P_n^{(N)}=|\sum_{m=0}^N U^{(N)}_{n,m} c^{(N)}_m|^2/\sum_{n=0}^N |\sum_{m=0}^N U^{(N)}_{n,m} c^{(N)}_m|^2$, where $U^{(N)}_{m,n}(z)=\bra{N-m,m}\hat{U}(z)\ket{N-n,n}$ are the matrix elements of the evolution operator $\hat{U}(z)=\exp(-iz\hat{H}^{(N)})$ in the $N$-photon subspace governed by the Hamiltonian $\hat{H}^{(N)}$  whose elements are $H^{(N)}_{m,n}=\braket{N-m,m|\hat{H}|N-n,n}$.
Remarkably, comparing the eigenmodes of $\hat{H}^{(N)}$, namely $\ket{\phi_E^{(N)}}=(1/\sqrt{2})^N \sqrt{N!}\sum_{n=0}^N i^{N-n}/\sqrt{n!(N-n)!}\ket{N-n}_{1}\ket{n}_{2}$, with $\ket{\psi^{(N)}}$ reveals that the coherent EM is composed of all $N$-photon eigenmodes, $\ket{i\alpha/\sqrt{2}}_{1}\ket{\alpha/\sqrt{2}}_{2}=\sum_{N=0}^\infty e^{-|\alpha|^2/2}\alpha^N/\sqrt{N!}\ket{\phi_E^{(N)}}$. From a practical point of view, this feature is very appealing since exact multiphoton eigenstates are extremely difficult to prepare and, as we show here, a coherent superposition of the N-photon eigenstates naturally appears in the coherent EM $\ket{\phi_E}$.\\
In Fig.~(\ref{fig:exceptionalmode}-b) we show the evolution of the photon number distribution computed from the amplitudes in the two waveguides, $P^{(N)}(z)=\sum_{n=0}^N|c_n^{(N)}|^2=e^{-|\alpha_1(z)|^2-|\alpha_2(z)|^2}(|\alpha_1(z)|^2+|\alpha_2(z)|^2)^N/N!$, for the particular case $\alpha=\sqrt{20}$.
In both, the critical and super-critical regime $P^{(N)}(z)$ describes a Poisson distribution with a decaying average photon number $\bar{n}(z)=|\alpha_1(z)|^2+|\alpha_2(z)|^2$. At the critical point (see solid line in Fig.~(\ref{fig:exceptionalmode}-b)) $\bar{n}_c(z)$ first decays more slowly than in the super-critical regime $\bar{n}_{sc}(z)$ (see dashed line Fig.~(\ref{fig:exceptionalmode}-b)). However, after the propagation distance $z_c$, which is the point where the solid and dashed lines cross, $z_c \approx 0.68 /\kappa$, the situation is reversed and we observe a higher average photon number in the super-critical regime, that is, the effect of loss-induced transparency. For example, at the specific distance $z_f=1/\kappa$, we find $\bar{n}_c(z_f)\approx 2.70$ and $\bar{n}_{sc}(z_f)\approx 4.02$ which corresponds to the classical increase in brightness by a factor of $\approx 1.5$ due to the second-order EP. \\
This situation is different when considering the post-selected $N$-photon subspace, as we show in Fig.~(\ref{fig:exceptionalmode}-c) for the specific photon number $N=6$. Initially, the distribution $P_n^{(6)}$ exhibits a symmetric, discrete Gaussian-like shape with highest probability $P_3^{(6)}\approx 30 \% $ for the state $\ket{3,3}$. Since the underlying $6$-photon EM $\ket{\phi_E^{(6)}}$ is an eigenstate of the 6-photon Hamiltonian, we would observe a global damping of all probabilities. However, this is cancelled exactly by the normalization requirement of probability distributions, leading to a distribution at the critical point which remains constant during propagation. On the other hand - in the super-critical regime - $\ket{\phi^{(6)}_E}$ is no longer an eigenstate of the 6-photon Hamiltonian and we expect an uneven damping of the amplitudes $c_n^{(N)}(z)$, which we indeed observe as more and more photons are detected in the neutral waveguide during propagation. For example, for the state $\ket{0,6}$ we initially observe a probability of $P_6^{(6)}(0/\kappa)\approx 1.5 \%$, which due to the EP of 7th order, is enhanced dramatically  by a factor of about $44$ to $P_6^{(6)}(2.5/\kappa)\approx 66.0 \%$.\\
\begin{figure}[t!]
\centering
\includegraphics[width=1\linewidth]{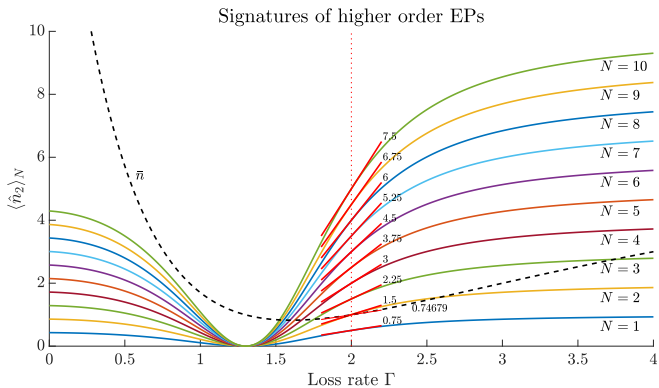}
\caption{Sensitivity comparison between classical (dashed black line) and quantum-post-selected sensing schemes (colored solid lines). The solid lines correspond to $\braket{\hat{n}_2}_N$ - the average number of photons in the neutral waveguide, given a total number of detected photons $N$ - while the dashed curve corresponds to the average total number of photons $\bar{n}$. In both cases, we have fixed the coupling coefficient $\kappa=1$, the propagation distance $z_f=1.5$ and the initial amplitude to $\alpha=\sqrt{20}$. For varying values of the loss-parameter $\Gamma$ we have then calculated the output-state and the corresponding observables. To indicate the sensitivity for the detection of a change in $\Gamma$, we have added red lines, accompanied by the numerically calculated slope at the working point $\Gamma=2$ (critical point) of the system. The slope corresponding to the quantum-post-selected scheme scales linearly with $N$ due to the order $N+1$ of the corresponding exceptional point of the system in the $N$-photon subspace.}
\label{fig:sensing}
\end{figure}
These results suggest that this device may find application in sensing. To illustrate this, consider the scenario where the propagation distance $z_f$ is fixed. To ensure that we are always in the regime of loss-induced transparency, we have to choose $z_f>z_c$, which is achieved by fixing $z_f=1.5/\kappa$. Further, we choose the ratio between $\kappa$ and $\Gamma$, such that the system operates near the critical point $\Gamma=\Gamma_c$. Under excitation with the coherent EM $\ket{\phi_E}$, any environmentally-induced losses $\Gamma \rightarrow \Gamma_c+\Delta \Gamma$ will drive the system into the super-critical regime. Then, using two PNR detectors at the output, one will observe - firstly, an increase in the average total number of photons as a function of the loss-parameter $\bar{n}(\Gamma)$, which will be governed by a 2nd order EP - and secondly, a shift in the post-selected $N$-photon distributions towards the neutral waveguide, which will be governed by a $(N+1)$'th order EP. To quantify the latter, a suitable observable is the photon number operator $\hat{n}_2$ in the neutral waveguide, which when applied to the $N$-photon state $\ket{\psi^{(N)}}=\sum_{n=0}^N c_n^{(N)} \ket{N-n,n}$, renders the average $\braket{\hat{n}_2}_N=\sum_{n=0}^N n |c_n^{(N)}|^2 \in [0,N]$. 
In Fig.~(\ref{fig:sensing}) we show the dependence of the observables $\braket{\hat{n}_2}_N$ and $\bar{n}$ on the loss-rate $\Gamma$, where the red lines indicate the slope at the critical point of the system. For the chosen parameters, the post-selected scheme outperforms the classical one. However, we must stress that the slope of $\bar{n}$ at the working point can be made arbitrarily large, simply by increasing the input power $\alpha$ of the coherent EM $\ket{i\alpha,\alpha}$, while the input-power has no effect on the post-selected distributions. This post-selected higher-order EP scheme is expected to yield improvements over classical schemes in the low-power regime, and more importantly, such higher order EPs can be observed and exploited using presently available technology. 

\begin{figure*}[t!]
\includegraphics[width=\linewidth]{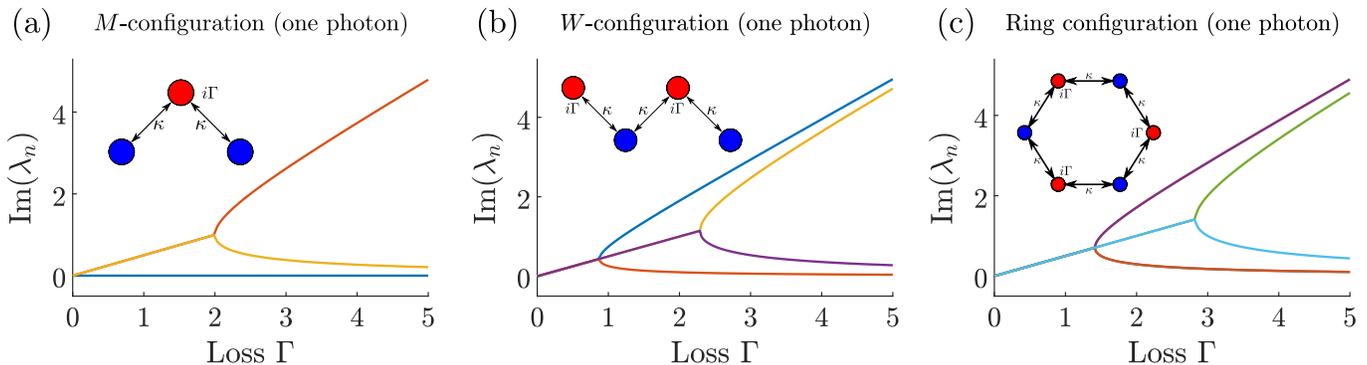}
\caption{Single photon exceptional points of partially lossy linear arrays of coupled waveguides exhibiting an alternating pattern of lossy/lossless waveguides. \textbf{(a)} A waveguide trimer ($\kappa=1/\sqrt{2}$) with a lossy central waveguide produces an EP of second order at $\Gamma_c=2$. \textbf{(b)} The waveguide tetramer ($\kappa=1/\sqrt{2}$) with two dissipative waveguides exhibits two exceptional points at $\Gamma_{c_\pm}=\sqrt{6\pm 2 \sqrt{5}}/\sqrt{2}$. \textbf{(c)} Ring configuration of 6 waveguides excited by a single photon.}
\label{fig:singlephotonEPs}
\end{figure*}
We now turn our attention to show how this scheme enables the study of higher order exceptional points with more exotic structures. To this end, we generalize the proposed scheme to systems of $M$ waveguides and identify realizations of $\hat{H}$ which display one, or possibly more, second order EPs. 
In general, one obtains EPs when the underlying waveguide structure exhibits a pattern of lossy and lossless waveguides which is symmetric under the parity operator - i.e. $\mathcal{P}\Gamma_m=\Gamma_{M+1-m}=\Gamma_m$ (up to a cyclic relabeling of the waveguides). Here, we focus on alternating patterns in $M$-configurations with $\Gamma_{2m+1}=0, \Gamma_{2m}=\Gamma$, $W$-configurations with 
$\Gamma_{2m+1}=\Gamma, \Gamma_{2m}=0$, and for systems with periodic boundary conditions.
We stress that for the emergence of EPs it is not necessary to include equal loss rates $\Gamma$, however, doing so greatly simplifies the analysis. In Fig.~(\ref{fig:singlephotonEPs}) we show three examples of such partially lossy coupled structures, a waveguide trimer $M=3$ ($M$-configuration), tetramer $M=4$ ($W$-configuration) and a waveguide hexagon (periodic) with one, two and three lossy waveguides, respectively. The plots illustrate how the imaginary part of the eigenvalues of the corresponding Hamiltonian vary as a function of the dissipation coefficient $\Gamma$. While the waveguide trimer features one EP at $\Gamma_c=2\sqrt{2}\kappa$, we observe two EPs $\Gamma_{c,\pm} = \sqrt{6\pm2\sqrt{5}}\kappa$ in the tetramer configuration, which are all of second order. And, the hexagonal waveguide configuration displays two second order EPs.\par

\begin{figure*}[t!]
\includegraphics[width=\linewidth]{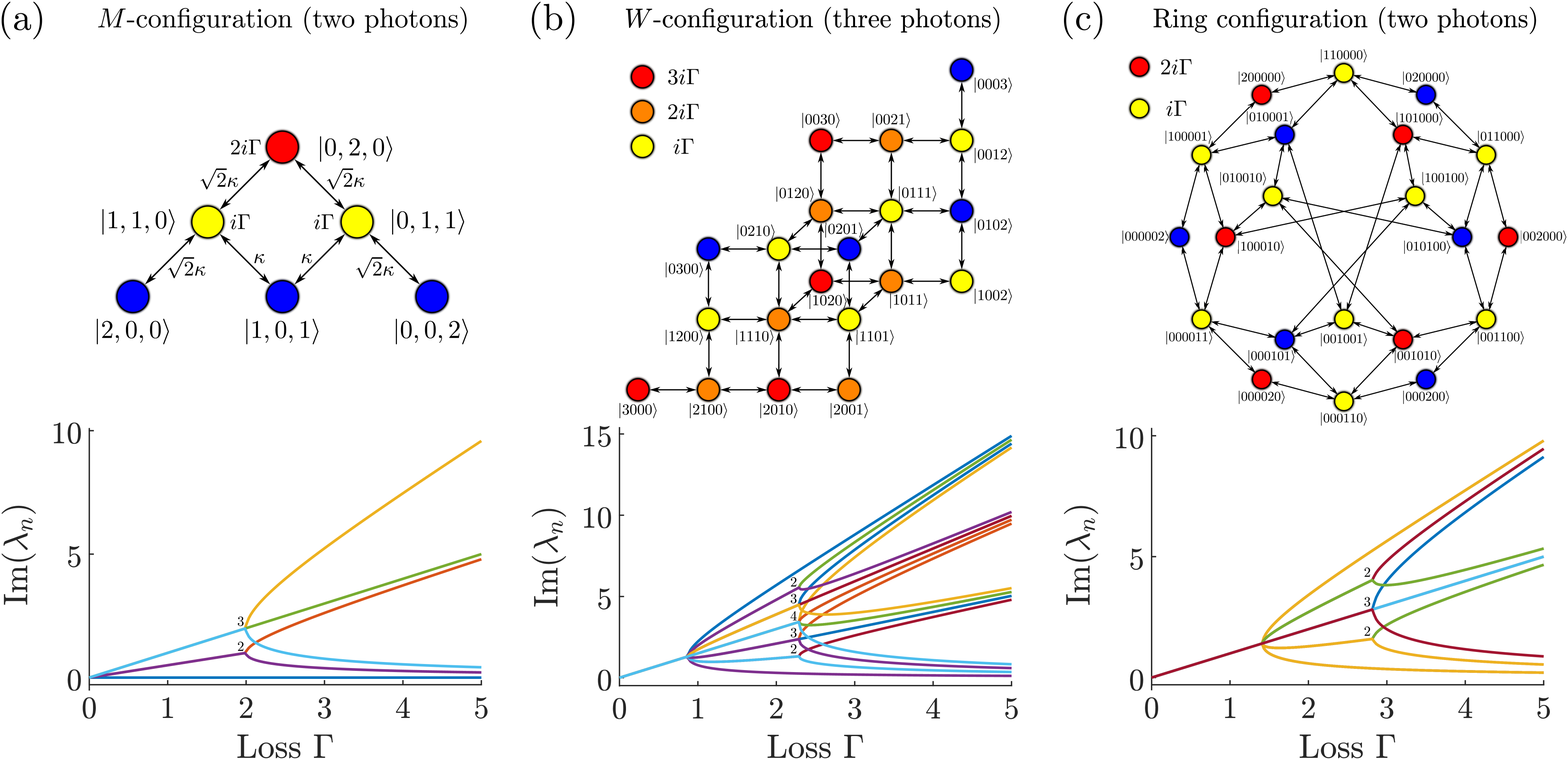}
\caption{Branching exceptional points of partially lossy waveguide arrays excited by two and three photons. \textbf{(a)} The balanced waveguide trimer ($\kappa=1/\sqrt{2}$) with a lossy central waveguide excited by two photons exhibits two EPs of second and third order at $\Gamma_c=2$. The order is  indicated by the numbers next to the bifurcation points. Top panel of \textbf{(a)} shows the synthetic two-photon structure in Fock space which is equivalent to a coupled system with 6 sites, three of which are dissipative. \textbf{(b)} Balanced waveguide tetramer ($\kappa=1/\sqrt{2}$) with two dissipative waveguides and excited by three photons. In this case the Fock graph becomes non-planar and exhibits multiple multifurcations at the dissipation rates $\Gamma_{c_\pm}=\sqrt{6\pm 2 \sqrt{5}}/\sqrt{2}$. \textbf{(c)} Top panel shows the synthetic coupled structure arising from the two photon excitation of a ring configuration of 6 waveguides. Bottom panel shows the spectrum versus the loss term.}
\label{fig:multiphotonEPs}
\end{figure*}

We now excite these systems with the coherent EM $\ket{\phi_E}$ corresponding to any of the critical points $\Gamma_c$. In general, we can express $\ket{\phi_E}=\ket{c_1 \alpha, c_2 \alpha, \ldots, c_M \alpha}$, with normalized amplitudes $\sum_{m=1}^M|c_m|^2=1$. Then, using $M$ photon-number-resolving detectors, we can project $\ket{\phi_E}=\sum_{N=0}^\infty \ket{\psi^{(N)}}$ onto the $N$-photon subspace $\ket{\psi^{(N)}}=\sum_{n_1+\ldots+n_M=N}$ $\braket{n_1,\ldots,n_M|\phi_E}\ket{n_1,\ldots,n_M}$, where the sum runs over all combinations of positive integers $0 \leq n_1,\ldots,n_M \leq N$, such that their sum yields $N$. In short, we have $\ket{\psi^{(N)}}=e^{-|\alpha|^2/2}\alpha^N \sum_{n_1+\ldots+n_M=N} \prod_{m=1}^M (c_m)^{n_m}/\sqrt{n_m!}\ket{n_1,\ldots,n_M}$, while the corresponding $N$-photon-$M$-mode EM can be expressed as $\ket{\phi_E^{(N)}}=\sqrt{N!}\sum_{n_1+\ldots+n_M=N} \prod_{m=1}^M$ $(c_m)^{n_m}/\sqrt{n_m!}\ket{n_1,\ldots,n_M}$. As in the two port case, also here the coherent $M$-mode EM is an infinite superposition of all $N$-photon-$M$-mode EM's $\ket{\phi_E}=\sum_{N=0}^\infty e^{-|\alpha|^2/2} \alpha^N/\sqrt{N!}\ket{\phi_E^{(N)}}$. Therefore, the PNR-detection of the output state will enable us to study the dynamics around higher order EPs in highly multi-mode configurations.\\
To illustrate the photon dynamics in the projected $N$-photon subspace, in Fig.~(\ref{fig:multiphotonEPs}) we depict the effective $N$-photon Hamiltonians as Fock graphs \cite{Tschernig:20}, where the nodes represent the Fock states $\ket{n_1,\ldots,n_M}$ with $\sum_{m=1}^M n_m =N$. The edges indicate allowed transitions, where a single photon tunnels between adjacent waveguides, i.e. $\ket{n_1,\ldots,n_m,n_{m+1},\ldots,n_M}\leftrightarrow\ket{n_1,\ldots,n_m\pm 1,n_{m+1}\mp 1,\ldots,n_M}$, with the corresponding effective coupling rate $\kappa \sqrt{(n_m+1)n_{m+1}}$ or $\kappa \sqrt{n_m(n_{m+1}+1)}$, respectively. Crucially, the effective loss-rates $\Gamma_{n_1,\ldots,n_M}=\sum_{m=1}^M n_m \Gamma_m$ are now enhanced by the number of photons occupying the lossy waveguides.

\begin{figure}[t!]
\includegraphics[width=\linewidth]{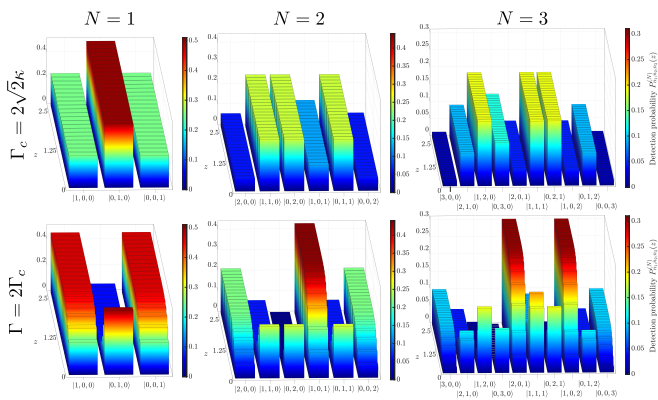}
\caption{Evolution of the post-selected $N$-photon EMs in the waveguide trimer, where the central waveguide is lossy. The columns correspond to the cases $N=1,2,3$, respectively. In the top row the system is at the critical point $\Gamma=2\sqrt{2}\kappa$ and in the bottom row it is in the super-critical regime $\Gamma=2\Gamma_c$. At the critical point, all states are proper eigenstates of the system, while in the super-critical regime more and more photons are detected in the neutral (outer) waveguides.}
\label{fig:3waveguides}
\end{figure}

For instance, in Fig.~(\ref{fig:multiphotonEPs}-a) there are three Fock states, $\ket{1,1,0},\ket{0,2,0}$ and $\ket{0,1,1}$, which are subjected to dissipation and $\ket{0,2,0}$ experiences twice the losses, since both photons reside in the dissipative waveguide. Further, since the photons do not interact, the eigenvalues $\lambda^{(N)}_{n_1,\ldots,n_M}=\sum_{m=1}^M n_m \lambda_m$ of the Hamiltonian in the $N$-photon subspace are given by sums of the single-photon eigenvalues $\lambda_m$ \cite{Tschernig:20}. This leads to an enhanced degeneracy of the $N$-photon eigenvalues and a large number of coalescent $N$-photon eigenmodes, which ultimately gives rise to higher order exceptional points as illustrated in the bottom panels of Fig.~(\ref{fig:multiphotonEPs}). 
Here, the increased order of the EPs is evidenced by the emergence of multiple multifurcation branches at different critical dissipation rates. To be precise, a system of $M$ waveguides will exhibit $\small\begin{pmatrix}N+M-2\\N\end{pmatrix}\normalsize$ multifurcation points in the $N$ photon subspace, while in the classical case it will exhibit a 2nd order EP with two coalescent eigenmodes $\hat{\phi}_1^\dagger=\hat{\phi}_2^\dagger$. Note, $\hat{\phi}_m^\dagger$ is the single-photon creation operator in the m-th eigenmode. 
In general, there will be $\small\begin{pmatrix} N-N_E+M-3 \\ N-N_E \end{pmatrix}\normalsize$ multifurcation points exhibiting $N_E+1$ branches, where  $N_E=0,\ldots,N$ represents the number of photons populating the coalescent single photon modes. As an example, let us denote the $N$-photon-$M$-mode eigenstates in the Fock-representation as $\ket{n_1,\ldots,n_M}=\prod_{m=1}^M (\hat{\phi}^\dagger_m)^{n_m}/\sqrt{n_m!}\ket{0}$. With this, we observe in the most extreme case $N_E=N$ that the set of modes $\{ \ket{N,0,0,\ldots,0} =\ket{N-1,1,0,\ldots,0}=\ldots$ $= \ket{0,N,0,\ldots,0} \}$ coalesce into the same mode, thus forming an EP of order $N+1$. However, also lower order EPs - i.e. sets with a smaller number of coalescent $N$-photon modes - will be observed at the same critical dissipation rate for $N_E<N$. For the sake of clarity, we have added numbers at the eigenvalue multifurcations to indicate the number of coalescent eigenmodes in Fig.~(\ref{fig:multiphotonEPs}). \par
For further illustration, in Fig.~(\ref{fig:3waveguides}) we show the evolution of the post-selected $N$-photon distributions in the partially lossy waveguide trimer. Again, at the critical point the coherent EM produces proper $N$-photon eigenstates of the system. Beyond the critical point we observe that more and more photons will be detected in the outer neutral waveguides. \par
It is important to point out that the synthetic coupled structures in Fig.~(\ref{fig:multiphotonEPs}) cannot be reproduced in real space configurations of waveguides, since the additional cross-talk between adjacent waveguides would inhibit the emergence of the exceptional points. Furthermore, the experimental realization of these configurations would require the adjustment of only one loss rate for the trimer configuration and three loss rates for the hexagonal configuration. The excitation with the coherent exceptional mode produces - in principle - an infinite number of synthetic coupling structures with many lossy sites in parallel superposition, which are nonetheless perfectly tuned, as they are perfect integer multiples of the loss rate $\Gamma$ of the physical waveguides. This overcomes a major challenge encountered in previous attempts of engineering high-order EPs, since here the fine-tuning of the loss rates of many synthetic sites is achieved automatically by virtue of the indistinguishability of the photons \cite{Tschernig18}. This point is further emphasized in Fig.~(\ref{fig:multiphotonEPs}-b,c) where the arising Fock graphs are non-planar, i.e., they cannot be drawn in 2D without intersecting edges, and 
it is thus truly impossible to implement them for single photons in real space via coupled waveguides. Rather, they can only be realized in synthetic space via multiphoton excitations \cite{Tschernig:20}. Additionally, these synthetic structures feature a highly non-trivial distribution of high- and low-loss sites, which are nevertheless perfectly tuned to obtain the higher-order EPs. \par
Finally, we briefly comment on the preparation of the input coherent EM. In general, this can be achieved by prepending a state-preparation stage that is free of losses ($\Gamma_m=0$). As alluded to above, the coherent EM of the semi-lossy waveguide beamsplitter can be achieved by launching the state $\ket{\psi_0}=\ket{0,\alpha}$ into a 50/50 beam splitter. Similarly, the exceptional mode of the trimer configuration can be obtained by launching the state $\ket{\psi_0}=\ket{0}\ket{\alpha}\ket{0}$ into a lossless trimer after the propagation distance $z=\frac{\pi}{4\sqrt{2}\kappa}$. In the general case, a network comprising a finite number of beam splitters and phase shifters is sufficient to perform the transformation from a single-mode beam to the desired EM. This general method replaces the challenging task of the preparation of pure $N$-photon-$M$-mode Fock-states by the ease of the generation of coherent states. Nevertheless, to observe higher order EPs it is necessary to employ PNR detectors.\\

\section{Conclusion}
In this work we have shown that arrays of coupled waveguides exhibiting alternating patterns of lossy and lossless sites, gives rise to several EPs of second order at the single photon or, equivalently, the classical level. The excitation with the corresponding coherent EM promotes these EPs to an arbitrary order in the emerging synthetic coupled structure in the $N$-photon Fock-space, and this without the need of further fine-tuning of the loss-rates in the underlying physical system \cite{Furch:19}. Utilizing PNR detectors, it is then possible to exploit the effect of loss-induced transparency in the quantum optical regime. 
We have outlined this concept for the semi-lossy waveguide beamsplitter, which is excited by its coherent exceptional mode. Further, we generalized this scheme to an arbitrary number of waveguides to show that integrated quantum optics based on multiphoton states  is richer than its single-photon counterpart and it requires fewer waveguide elements as it works in synthetic space. The challenges of this approach is the need to generate multiphoton states and to have suitable PNR detectors. Using coherent states we greatly alleviate the problem of the generation of multiphoton states but it still relies on PNR detectors which is a manageable problem \cite{Lita08,Young2020,Zou2020}. Since our analysis is based on a tight-binding model, we anticipate that this scheme will find applications in numerous other formally equivalent systems, such as coupled micro-cavities, ring resonators or ultra-cold atoms trapped in an optical lattice.

%

\section*{Disclosures}
The authors declare no conflicts of interest.

\section*{Data availability}
No data were generated or analyzed in the presented research.

\bibliography{literature}

\end{document}